\pdfoutput=1
\documentclass[]{article}  
\usepackage{url,float}
\usepackage{graphicx}
\usepackage{amsmath}
\usepackage{amsfonts}
\usepackage{amssymb}
\usepackage{latexsym}

\newcommand{\hide}[1]{}

\newcommand{\ABox}{
\raisebox{3pt}{\framebox[6pt]{\rule{6pt}{0pt}}}
}
\newenvironment{proof}{{\bf Proof:}}{\hfill\ABox}

\newtheorem{theorem}{{\bf Theorem}}

\newtheorem{lemma}{Lemma}
\newtheorem{question}[lemma]{Question}

\newcommand{\lemlab}[1]{\label{lemma:#1}}
\newcommand{\thmlab}[1]{\label{thm:#1}}

\newcommand{\figlab}[1]{\label{fig:#1}}
\newcommand{\seclab}[1]{\label{sec:#1}}

\newcommand{\thmref}[1]{\ref{thm:#1}}
\newcommand{\secref}[1]{\ref{sec:#1}}
\newcommand{\figref}[1]{\ref{fig:#1}}

\def\R{{\mathbb{R}}}

\newcommand{\squeezelist}{\setlength{\itemsep}{0pt}}

\title{%
Hypercube Unfoldings that Tile $\R^3$ and $\R^2$
} 

\author{%
Giovanna Diaz%
    \thanks{Depts of Computer Science, and Mathematics, Smith College. 
      \protect\url{gdiaz@smith.edu}.}
\and
Joseph O'Rourke%
    \thanks{Depts of Computer Science, and Mathematics, Smith College, Northampton, MA 01063, USA.
      \protect\url{orourke@cs.smith.edu}.}
}

\begin{document}
\maketitle

\begin{abstract}
We show that the hypercube has a face-unfolding that tiles space,
and that unfolding has an edge-unfolding that tiles the plane.
So the hypercube is a ``dimension-descending tiler."
We also show that the hypercube cross unfolding made famous by Dali
tiles space, but we leave open the question of whether or not
it has an edge-unfolding that tiles the plane.
\end{abstract}


\section{Introduction}
\seclab{Introduction}
The cube in $\R^3$ has $11$ distinct (incongruent) 
edge-unfoldings\footnote{
An \emph{edge-unfolding} cuts along edges.
}
to $6$-square 
planar polyominoes, each of which tiles the plane~\cite{Etudes}.
A single tile (a \emph{prototile}) that tiles the plane with congruent copies
of that tile 
(i.e., tiles via translations and rotations, but not reflections)
is called a \emph{monohedral} tile.
The cube itself obviously tiles $\R^3$.
So the cube has the pleasing property that it tiles $\R^3$
and all of its edge-unfoldings tile $\R^2$.
The latter property makes the cube a \emph{semi-tile-maker}
in Akiyama's notation~\cite{JinAkiyama}, a property shared by
the regular octahedron.

In this note we begin to address a higher-dimensional analog of these
questions.
The 4D hypercube (or \emph{tesseract}) tiles $\R^4$.
Do all of its face-unfoldings monohedrally tile $\R^3$?
The hypercube  has $261$ distinct face-unfoldings
(cutting along $2$-dimensional square faces)
to $8$-cube polycubes,
first enumerated by Turney~\cite{Turney}
and recently constructed and confirmed by McClure~\cite{McClure_MO}~\cite{JOR_MO1}.
The second author posed the question of determining which
of the $261$ unfoldings tile space monohedrally~\cite{JOR_MO2}.

Whether or not it is even decidable to determine if a given tile can tile
the plane monohedrally is an open problem~\cite{JOR_TCS},
and equally open for $\R^3$.
The only general tool is Conway's sufficiency criteria~\cite{Schattschneider}
for planar prototiles,
which seem too specialized to help much here.
In the absence of an algorithm, this seems a daunting task.

Here we focus on two narrower questions, essentially replacing
Akiyama's ``all" with ``at least one":
\begin{question}
Is there an unfolding of the hypercube that tiles $\R^3$,
and which itself has an edge-unfolding that tiles $\R^2$?
\end{question}
Call a polytope that monohedrally tiles $\R^d$ 
a \emph{dimension-descending tiler} (DDT) if it 
has a facet-unfolding that tiles $\R^{d-1}$,
and that $\R^{d-1}$ polytope has a facet-unfolding that tiles $\R^{d-2}$,
and so on down to an edge-unfolding that tiles $\R^2$.
(Every polygon has a vertex-unfolding of its perimeter that 
trivially tiles $\R^1$.)
Thus the cube is a DDT. We answer Question~1 positively by
showing that the hypercube is a DDT, by finding
one face-unfolding to an $8$-cube polyform in $\R^3$, which
itself has an edge-unfolding to a $34$-square polyominoe that
tiles $\R^2$.

It is natural to wonder about the other $260$ face-unfoldings of the hypercube, and
in particular, the most ``famous" one,
what we call the \emph{Dali cross},
made famous in Salvadore Dali's painting shown in 
Figure~\figref{Dali_Crucifixion_hypercube}.
\begin{figure}[htbp]
\centering
\includegraphics[width=0.5\linewidth]{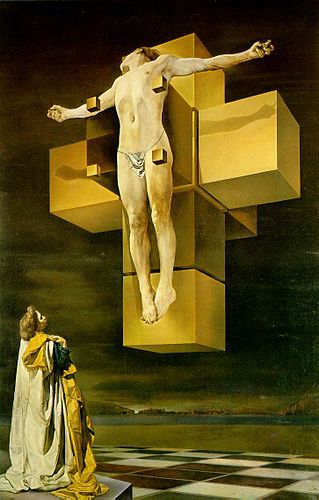}
\caption{The 1954 Dali painting \emph{Corpus Hypercubus}.
(Image from Wikipedia).}
\figlab{Dali_Crucifixion_hypercube}
\end{figure}

\begin{question}
Does the Dali cross tile $\R^3$, and if so, does it
have an edge-unfolding that tiles $\R^2$?
\end{question}
Here we are only partially successful: We show that the Dali cross
does indeed tile space (Theorem~\thmref{DaliCross}), but we have not succeeded in finding an unfolding
of this cross that tiles the plane.

\section{Hypercube Unfoldings that Tile $\R^3$}
\seclab{TileSpace}
So far as we are aware, there are now $4$ hypercube unfoldings that
are known to tile space.
The first two were found by Steven Stadnicki~\cite{Stadnicki_MO} in response
to the question raised in~\cite{JOR_MO2}.
We call the first of Stadnicki's unfoldings the $L$-unfolding.
We describe this in detail for it is the unfolding we use to answer Question~1.

\subsection{The Hypercube $L$-unfolding tiles $\R^3$}
\seclab{Unf-L}
The $L$-unfolding is shown in 
Figure~\figref{L1_3D}. (The labels will not be used until Section~\secref{EdgeUnf}.)
\begin{figure}[htbp]
\centering
\includegraphics[width=0.75\linewidth]{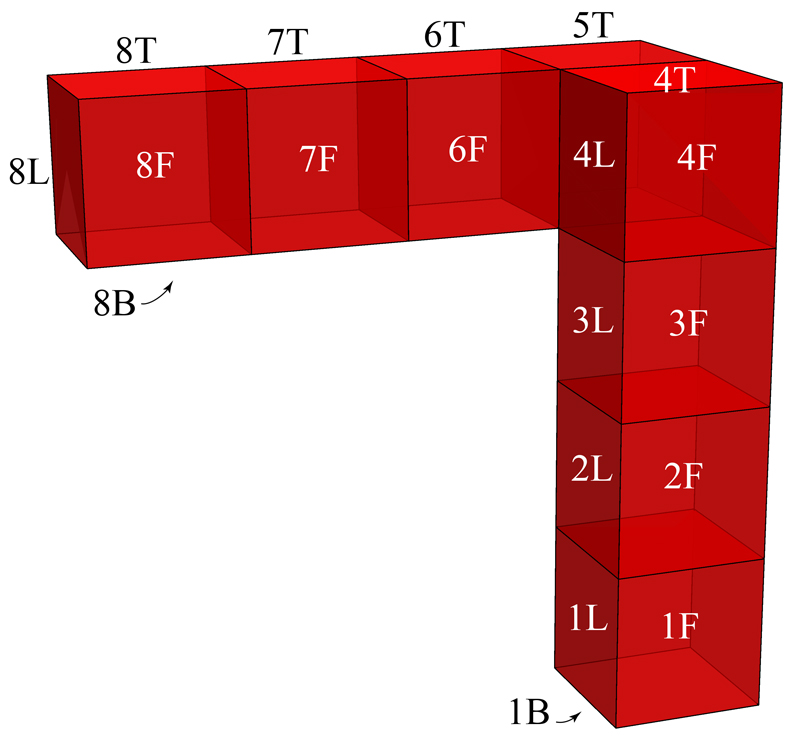}
\caption{The $L$-unfolding of the hypercube. Some face labels are shown.}
\figlab{L1_3D}
\end{figure}
Stadnicki showed this leads to a particularly simple tiling of space, because
nestling one $L$ inside another as shown in 
Figure~\figref{L5_3D}
\begin{figure}[htbp]
\centering
\includegraphics[width=0.75\linewidth]{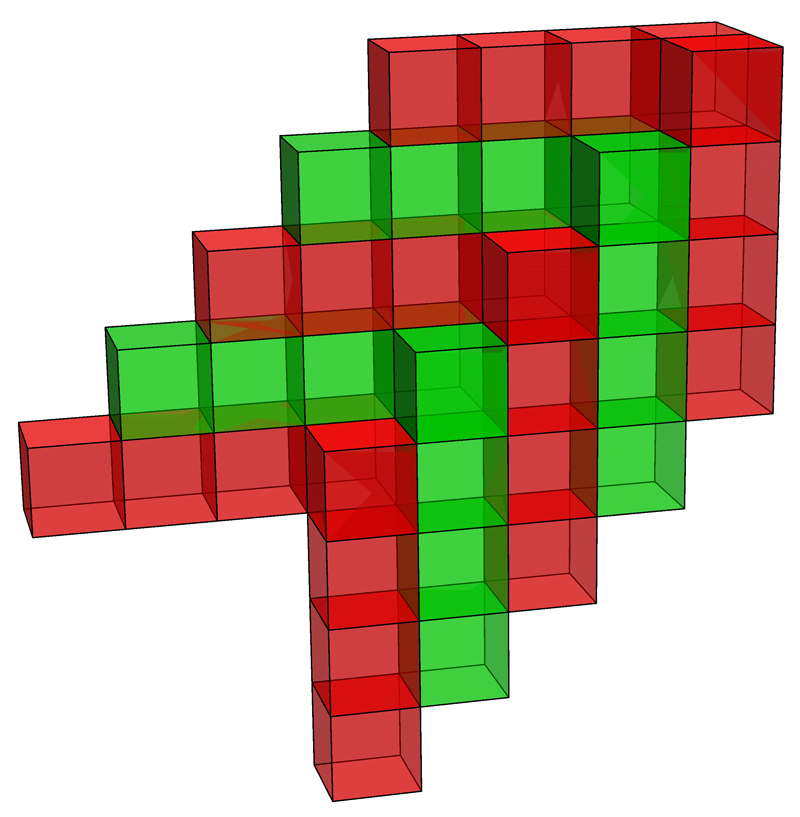}
\caption{Five nestled $L$'s.}
\figlab{L5_3D}
\end{figure}
leads to a $2$-cube thick infinite slab, as illustrated in
Figure~\figref{L10_3D}.
\begin{figure}[htbp]
\centering
\includegraphics[width=0.75\linewidth]{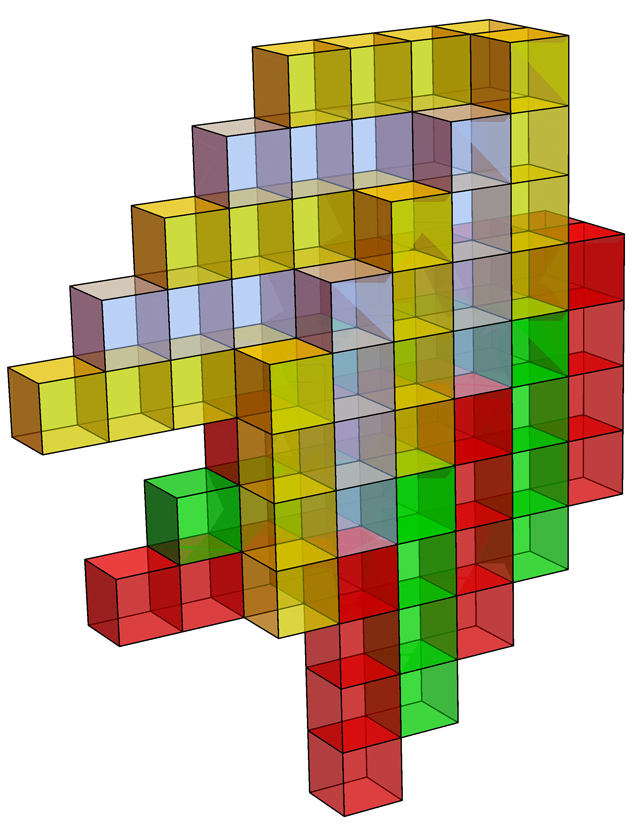}
\caption{Ten nestled $L$'s. Note the evolving structure is two-cubes thick in depth.}
\figlab{L10_3D}
\end{figure}
Then of course all of $\R^3$ can be tiled by stacking the $2$-cube thick slabs.
We will return to edge-unfolding the $L$ in Section~\seclab{EdgeUnf}.

Stadnicki showed that a second unfolding (Figure~\figref{Stadnicki})
also tiles space~\cite{Stadnicki_MO}, via a slightly more complicated but still simple
structure. We will not describe that tiling.
\begin{figure}[htbp]
\centering
\includegraphics[width=0.25\linewidth]{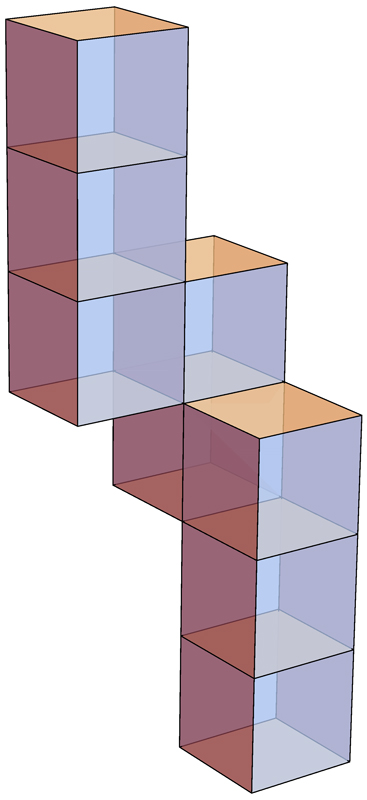}
\caption{Another hypercube unfolding that tiles $\R^3$ (Stadnicki).}
\figlab{Stadnicki}
\end{figure}

\newpage
\subsection{The Dali Cross Unfolding tiles $\R^3$}
\seclab{DaliCross}
Recall the Dali cross consists of four cubes in a tower, with the third tower-cube surrounded by four more;
see Figure~\figref{Cross1_3D}. 
(Again the labels will not be used until Section~\secref{EdgeUnf}.)
\begin{figure}[htbp]
\centering
\includegraphics[width=0.50\linewidth]{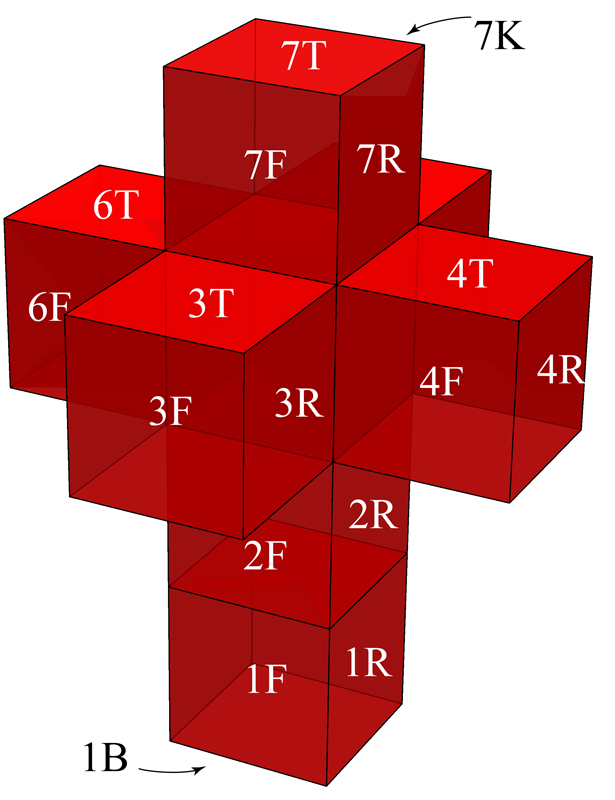}
\caption{The Dali cross. Some face labels are shown.}
\figlab{Cross1_3D}
\end{figure}

Our proof that this shape tiles $\R^3$ is in six steps:
\begin{enumerate}
\squeezelist
\item $2$-cross unit.
\item Cross-strip.
\item Cross-layer.
\item Two cross-layers.
\item Three cross-layers.
\item Four cross-layers.
\end{enumerate}

\subsection{$2$-Cross Unit}
We first build a $2$-cross unit with prone, opposing crosses, as illustrated in
Figure~\figref{Cross2_3D}.
We will call planes of possible cube locations $z$-layers $1, 2, 3, \ldots$,
corresponding to $z$-height.
The $2$-cross unit has two cubes in $z$-layers $1$ and $3$, in the same $xy$-locations,
and the remaining cubes in $z$-layer $2$.
It will be convenient to use \emph{bump} to indicate a cube protruding above
a particular layer of interest, and use \emph{hole} to indicate a cube cell 
as-yet unoccupied by a cube.
\begin{figure}[htbp]
\centering
\includegraphics[width=0.75\linewidth]{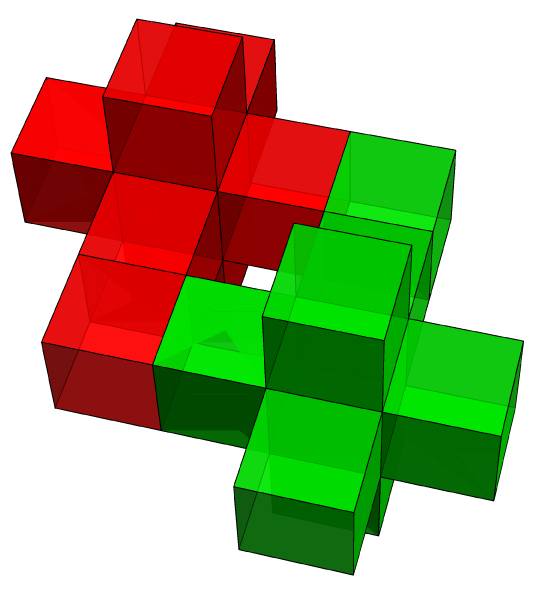}
\caption{A $2$-cross unit.}
\figlab{Cross2_3D}
\end{figure}

\subsection{Cross-strip}
Now we form a vertical strip of $2$-cross units
as shown in Figure~\figref{HC_1}.
Here we introduce a convention of displaying the construction by using colors
and $z$-layer numbers. So the cubes in a cross-strip occupy $z$-layers $1,2,3$, but
only $z$-layers $2$ and $3$ are visible from above in an overhead view.
\begin{figure}[htbp]
\centering
\includegraphics[width=0.75\linewidth]{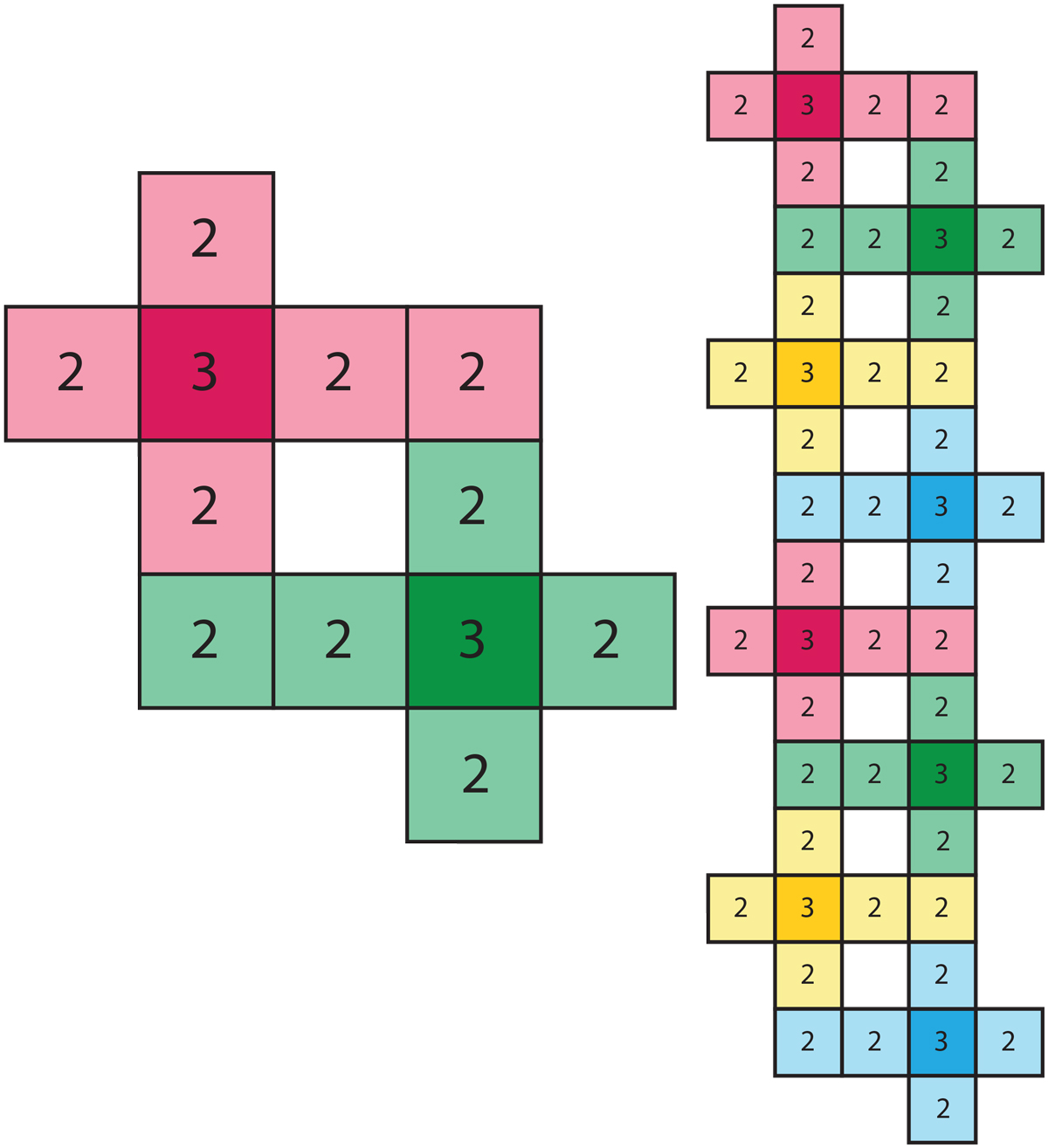}
\caption{Cross-strip.}
\figlab{HC_1}
\end{figure}

\subsection{Cross-layer}
Now we place cross-strips adjacent to one another horizontally,
as shown in Figure~\figref{HC_2}.
\begin{figure}[htbp]
\centering
\includegraphics[width=0.90\linewidth]{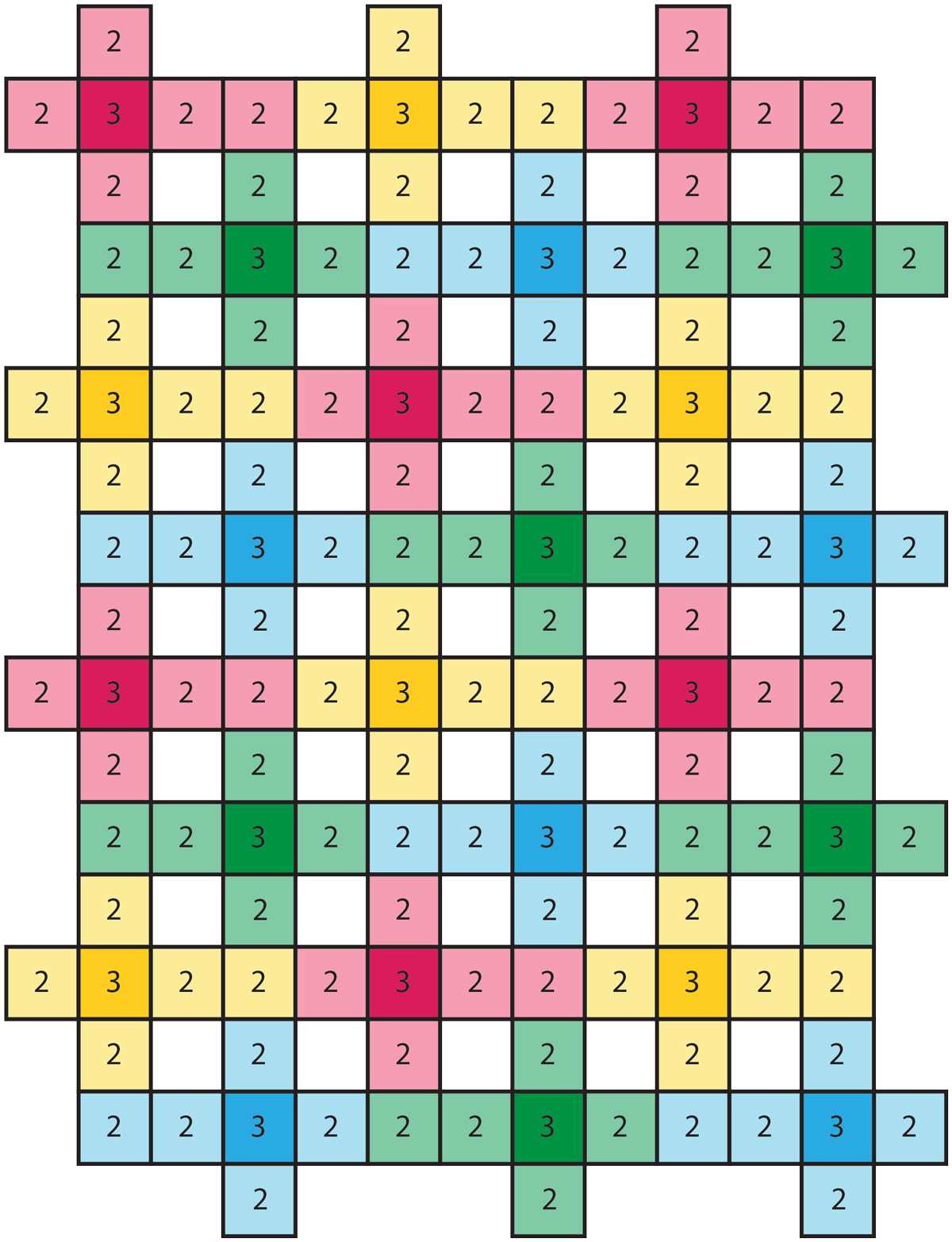}
\caption{Cross-layer.}
\figlab{HC_2}
\end{figure}
The remaining steps stack cross-layers one on top of the other.
So the pattern of holes and bumps in each cross-layer will be important.

\subsection{Two Cross-Layers}
Henceforth we color all cubes in one cross-layer the same primary color,
with the bumps slightly darker, as in Figure~\figref{HC_3}(a).
Remember the bumps in one
cross-layer align vertically.
Now we place a second cross-layer on top of the first,
with the bumps in the second cross-layer fitting into the holes of the
first.
Figure~\figref{HC_3}(b) shows the top view, which will be our focus.
Note that now we see cubes at $z$-layers $2,3,4$. That there are no holes
all the way through; rather, $z$-layer-$2$ cells are dents and $z$-layer-$4$ cells bumps.
\begin{figure}[htbp]
\centering
\includegraphics[width=1.00\linewidth]{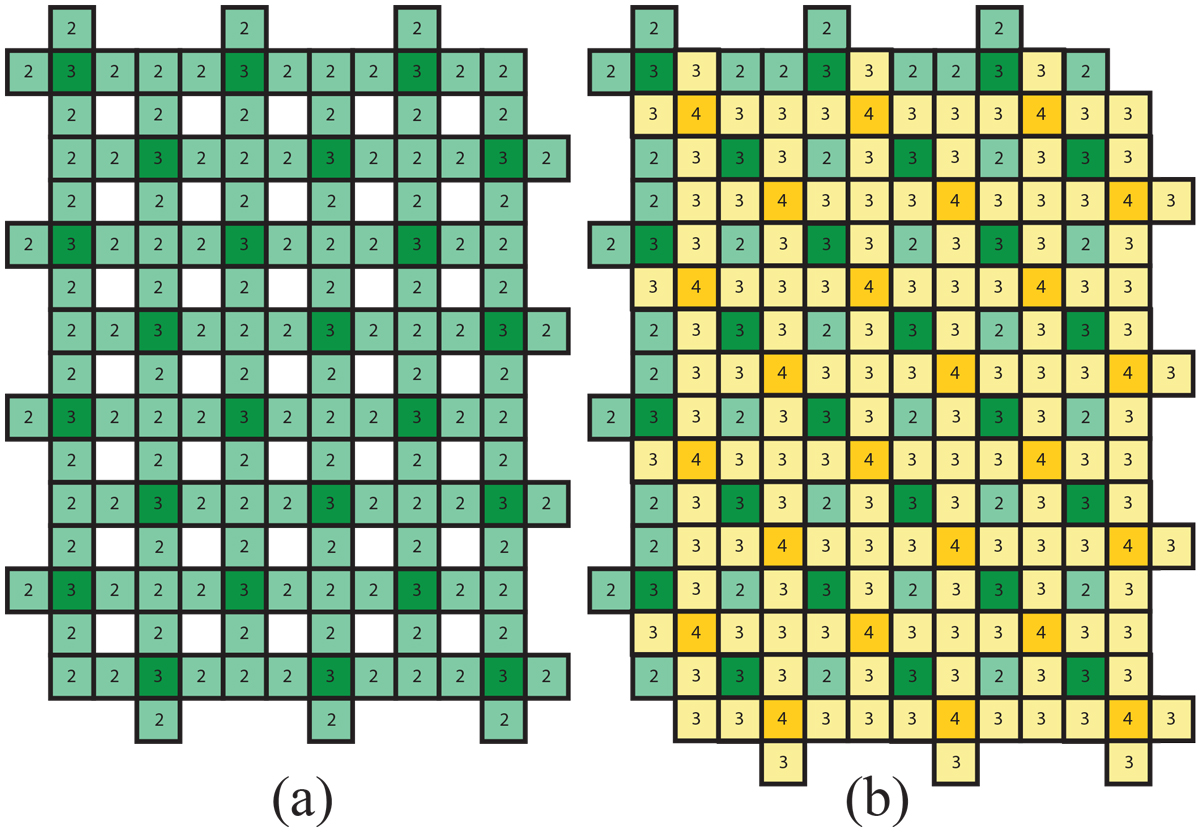}
\caption{Two cross-layers.
(a)~One cross-layer.
(b)~Two cross-layers.
}
\figlab{HC_3}
\end{figure}
We ask the reader to concentrate on the pattern depicted in Figure~\figref{HC_4}:
in two adjacent columns, we see $(4,3,3,3,4)$ and $(2,3,3,3,2)$, with the latter
pattern shifted diagonally upward one unit.
It should be clear that the entire overhead $z$-layer-view is composed of copies of this
\emph{fundmental layer-pattern}.
\begin{figure}[htbp]
\centering
\includegraphics[width=0.75\linewidth]{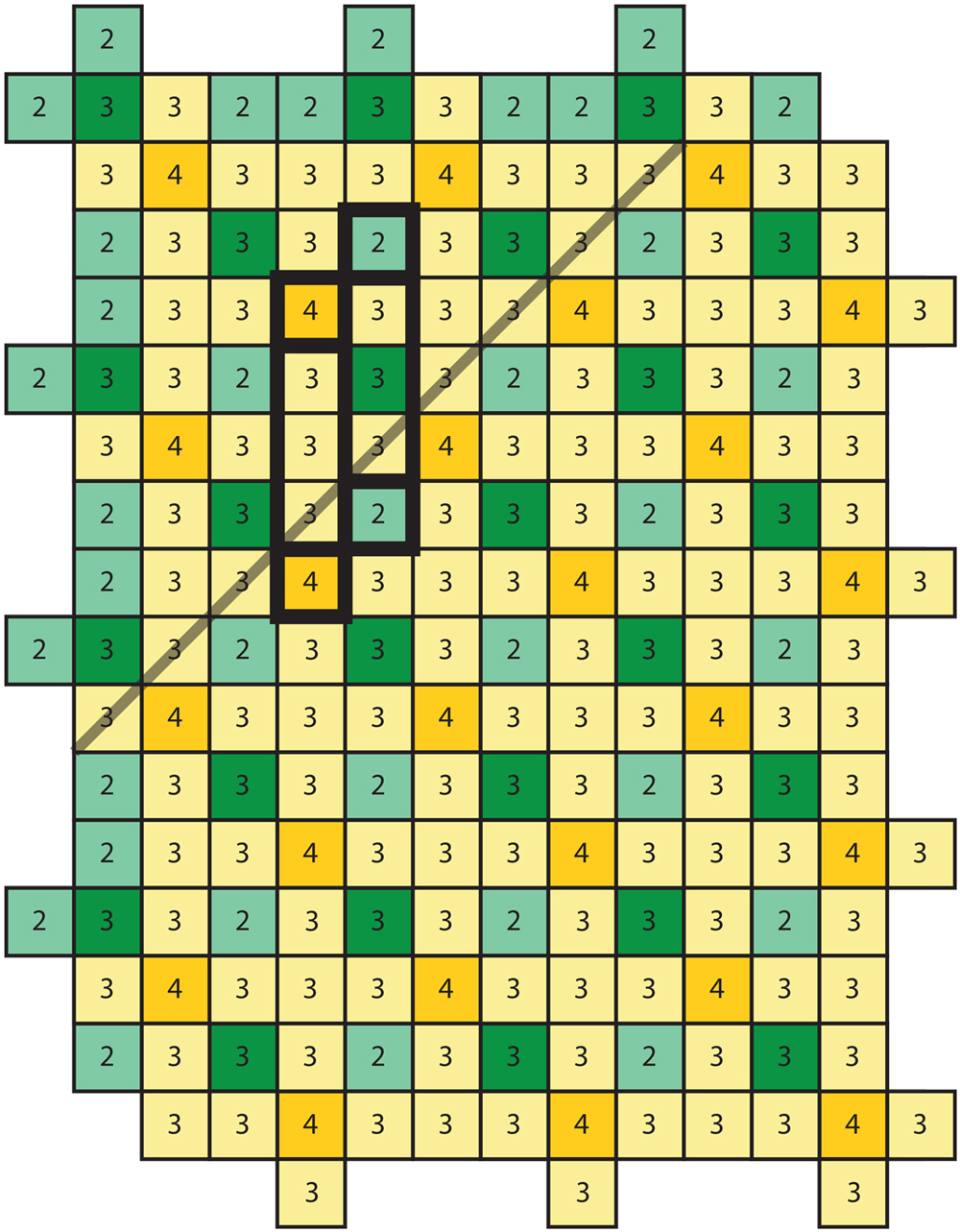}
\caption{Fundamental layer-pattern after stacking two cross-layers.}
\figlab{HC_4}
\end{figure}

\subsection{Three Cross-Layers}
When we stack a third cross-layer on the construction, again inserting
bumps into dents, we do not quite regain the fundamental layer-pattern.
Instead we see that pattern shifted diagonally downward rather than upward;
see Figure~\figref{HC_5}. Although we could argue that now we see
a reflection (over a horizontal) of the full pattern of visible $z$-layer numbers,
it seems easier and more convincing to us to add one more cross-layer.
\begin{figure}[htbp]
\centering
\includegraphics[width=0.75\linewidth]{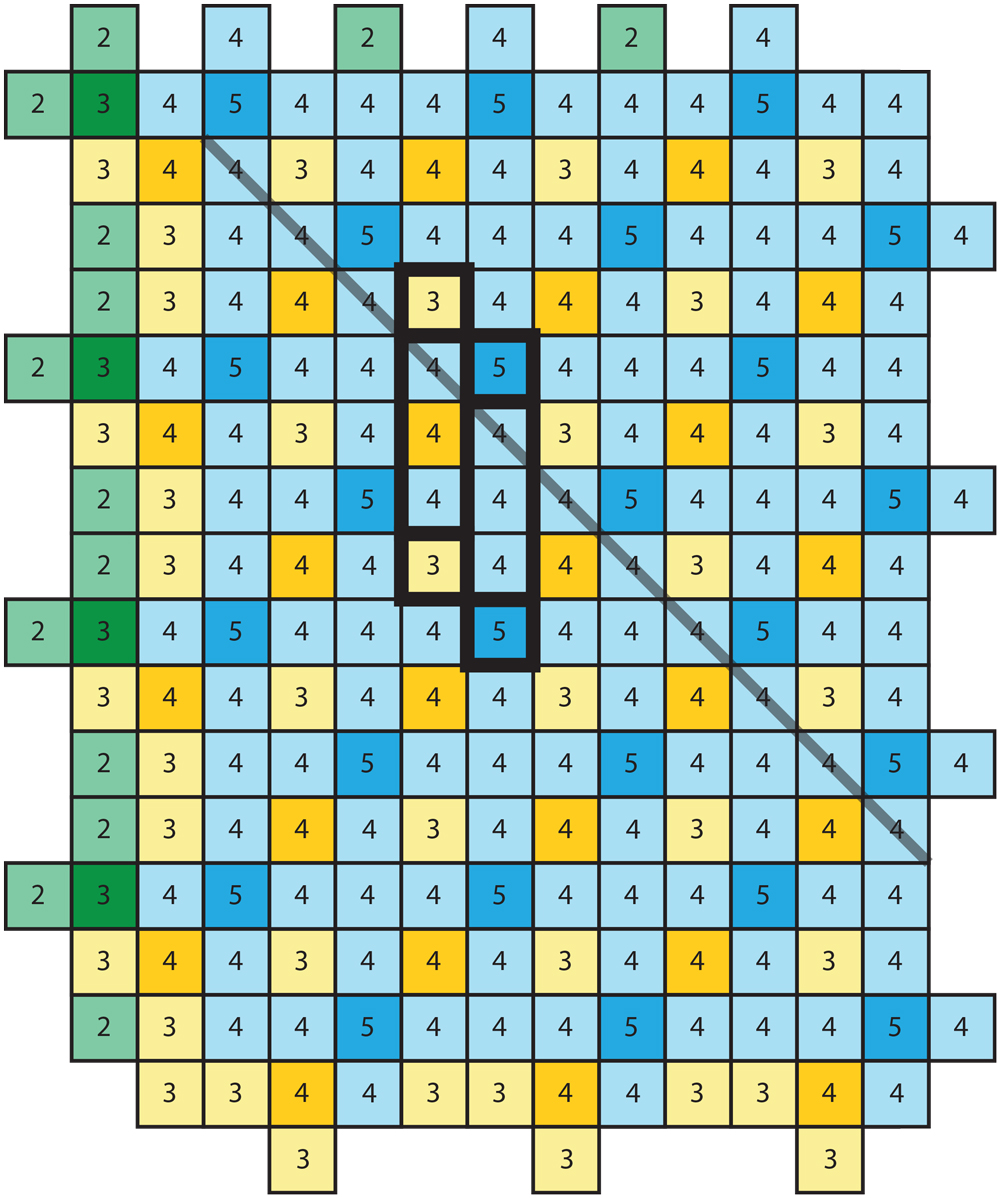}
\caption{Three cross-layers and a reflected pattern.}
\figlab{HC_5}
\end{figure}

\subsection{Four Cross-Layers}
With the addition of the fourth cross-layer (Figure~\figref{HC_6}),
we regain the exact same
pattern of $z$-layer numbers. Note the fundamental layer-pattern is
now $(6,5,5,5,6)$ and $(4,5,5,5,4)$, exactly $+2$ of the pattern in
two cross-layers, as emphasized in Figure~\figref{HC_7}.

\begin{figure}[htbp]
\centering
\includegraphics[width=0.75\linewidth]{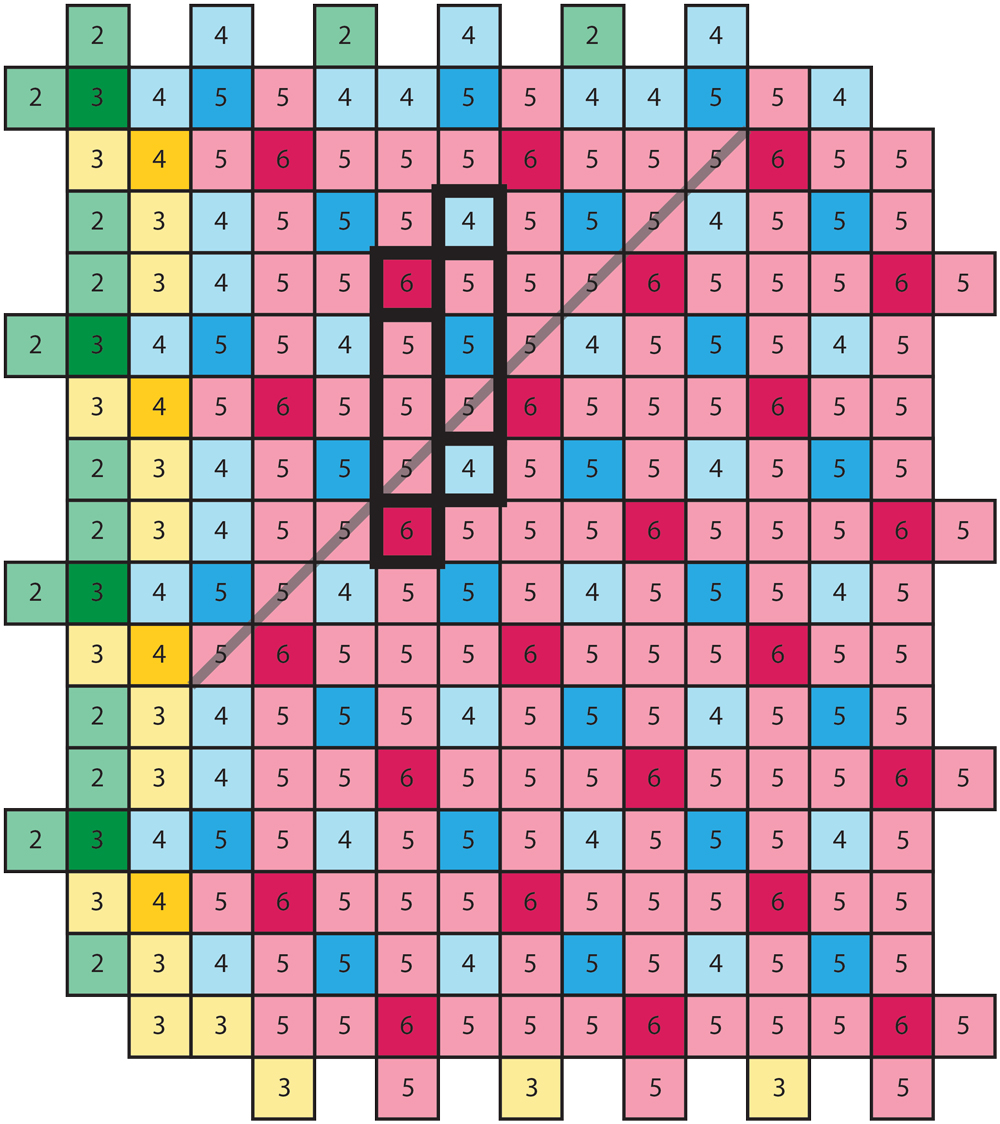}
\caption{Four cross-layers.}
\figlab{HC_6}
\end{figure}
\begin{figure}[htbp]
\centering
\includegraphics[width=1.00\linewidth]{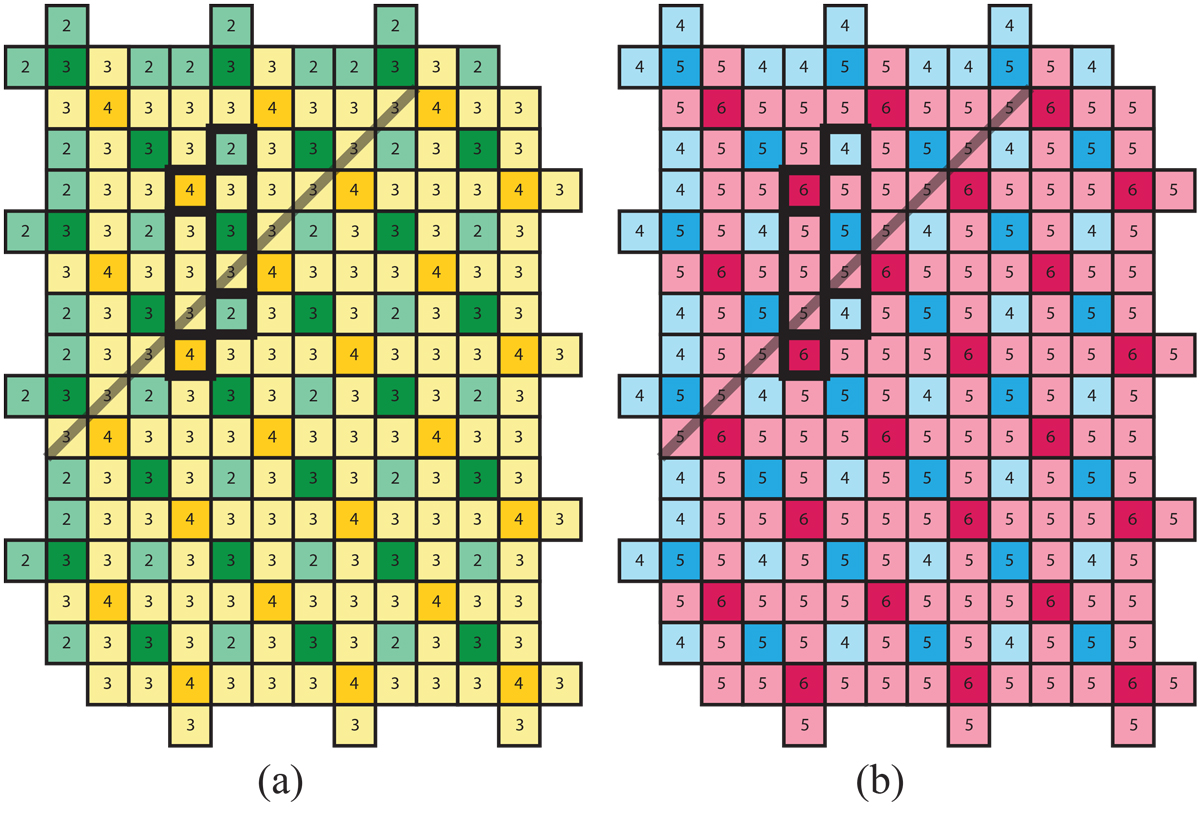}
\caption{Two cross-layers~(a) compared to four cross-layers~(b), with the same
fundamental pattern indicated.}
\figlab{HC_7}
\end{figure}

It is now clear that because we have regained at four cross-layers
the exact same ``$z$-layer landscape" as we had at two cross-layers,
the stacking can be continued indefinitely.
\begin{theorem}
The Dali cross unfolding of the hypercube tiles $\R^3$ monohedrally.
\thmlab{DaliCross}
\end{theorem}

We have found another hypercube unfolding, shown in Figure~\figref{TTunfolding},
that tiles $\R^3$ in a similar manner, not described here.
\begin{figure}[htbp]
\centering
\includegraphics[width=0.5\linewidth]{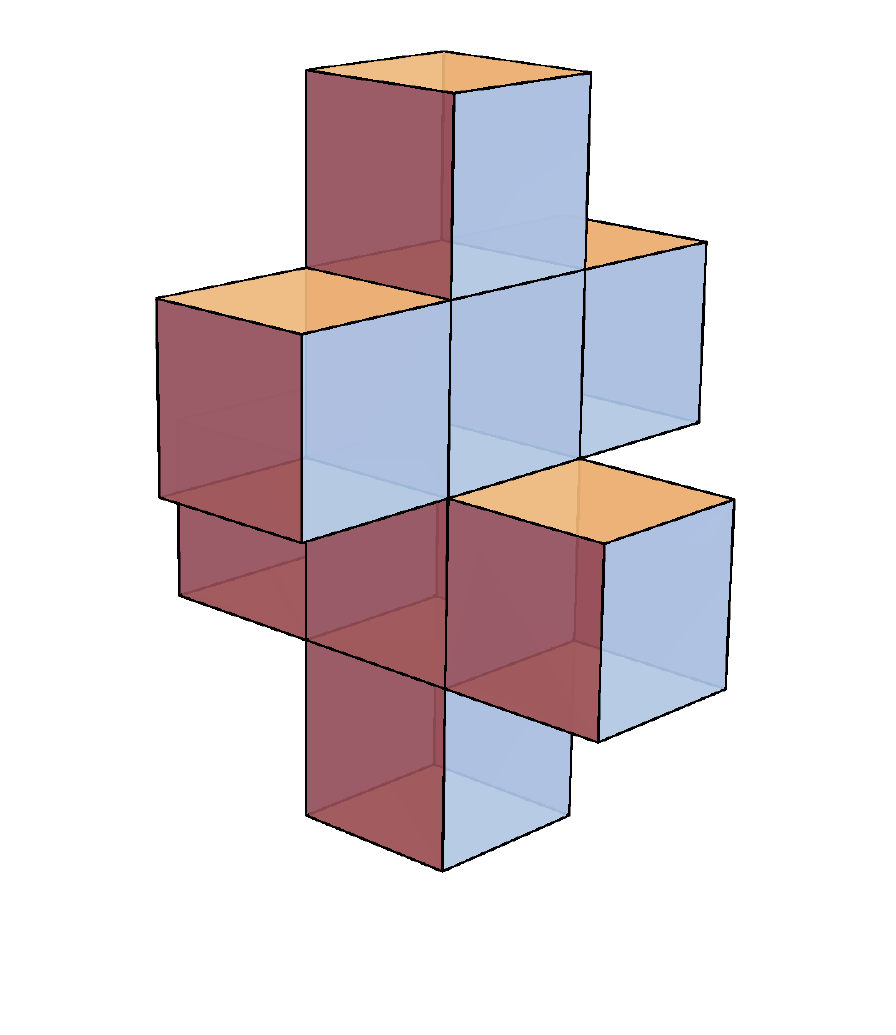}
\caption{Another hypercube unfolding that tiles $\R^3$.}
\figlab{TTunfolding}
\end{figure}

\newpage
\section{Edge-unfoldings to tile $\R^2$}
\seclab{EdgeUnf}

Now we turn to unfolding the $L$ to tile the plane.
We label the cubes from $1$ to $8$, and the faces
as $\{F, L, K, R, B, T\}$ for \{Front, Left, bacK, Right, Bottom, Top\} respectively.
Refer again to Figure~\figref{L1_3D}.
There are $34$ exposed faces of the $8$ cubes.
Through a mixture of heuristic computer searches and hand tinkering, we found
the unfolding shown in  
Figure~\figref{L_unfolding}.
\begin{figure}[htbp]
\centering
\includegraphics[width=0.5\linewidth]{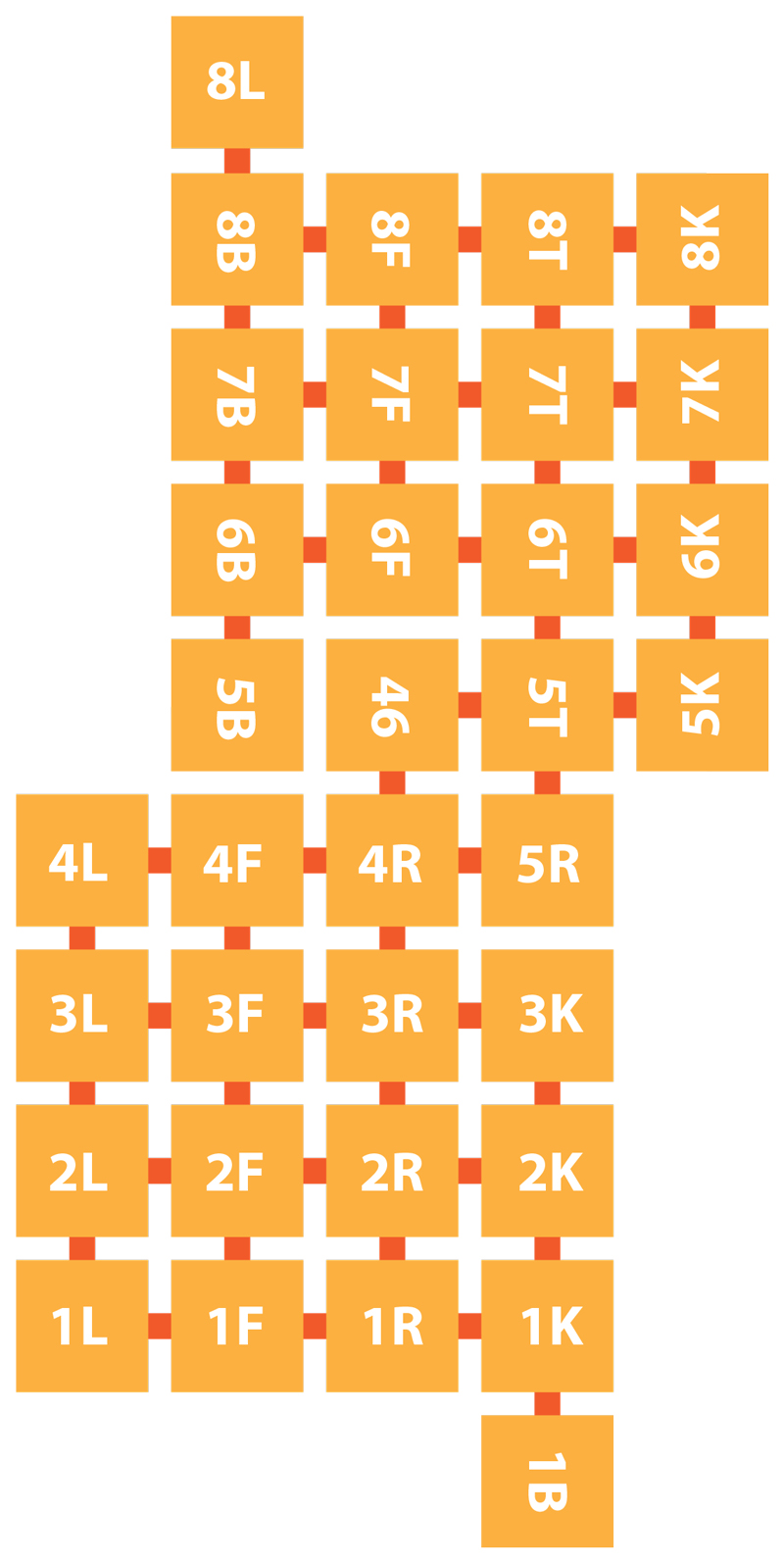}
\caption{Unfolding of the $L$ (Figure~\protect\figref{L1_3D}), 
with face labels and dual-tree (uncut) connections.}
\figlab{L_unfolding}
\end{figure}

That this tiles the plane (by translation only) is demonstrated
in Figure~\figref{tiledL_all}.
\begin{figure}[htbp]
\centering
\includegraphics[width=0.75\linewidth]{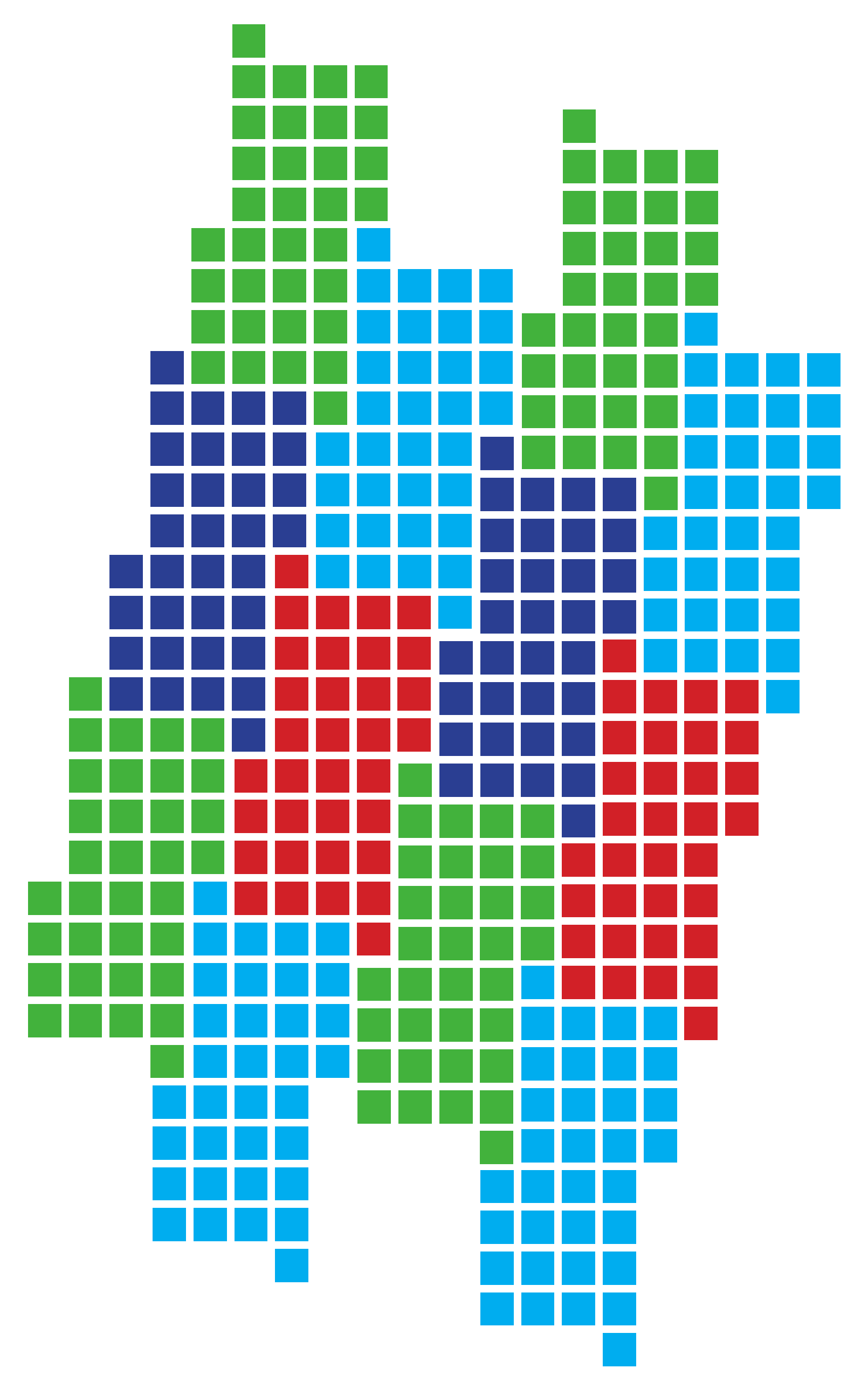}
\caption{Tiling of the plane with the unfolding shown in Figure~\protect\figref{L_unfolding}.}
\figlab{tiledL_all}
\end{figure}
This then establishes our answer to Question~1:
\begin{theorem}
The $L$-unfolding of the hypercube has an edge-unfolding
that tiles the plane, establishing that the hypercube is a
dimension-descending tiler.
\thmlab{LTilesPlane}
\end{theorem}

\subsection{Edge-unfoldings of the Dali cross}
\seclab{EdgeUnfDaliCross}
There are a huge number of edge-unfoldings of each hypercube unfolding.
Each edge-unfolding corresponds to a spanning tree of the dual graph, where
each square face is a node, and arcs represent uncut edges.
There are at most approximately $5^n$ spanning trees~\cite{GRote} 
of planar graphs with
$n$ nodes, and asymptotically that many for some graphs.
It seems conservative to estimate that the dual graph of the Dali cross has at least
$2^n=2^{34} \approx 10^{10}$ spanning trees,
and more likely $3^{34} \approx 10^{16}$.
(The square grid has $3.2^n$ spanning trees, and each hypercube unfolding
dual graph is also regular of degree $4$.)
Each of these trees leads to an unfolding, but many self-overlap in their
planar layout, and even among those that avoid overlap, many delimit
a region with holes, and so could not form tilers.
With brute-force search infeasible, and no algorithm available,
we are left only with heuristics, with which we have not been successful.

Figure~\figref{hypercrossUnfolding} shows the closest to a tiling
unfolding of the Dali cross that we found.
\begin{figure}[htbp]
\centering
\includegraphics[width=0.75\linewidth]{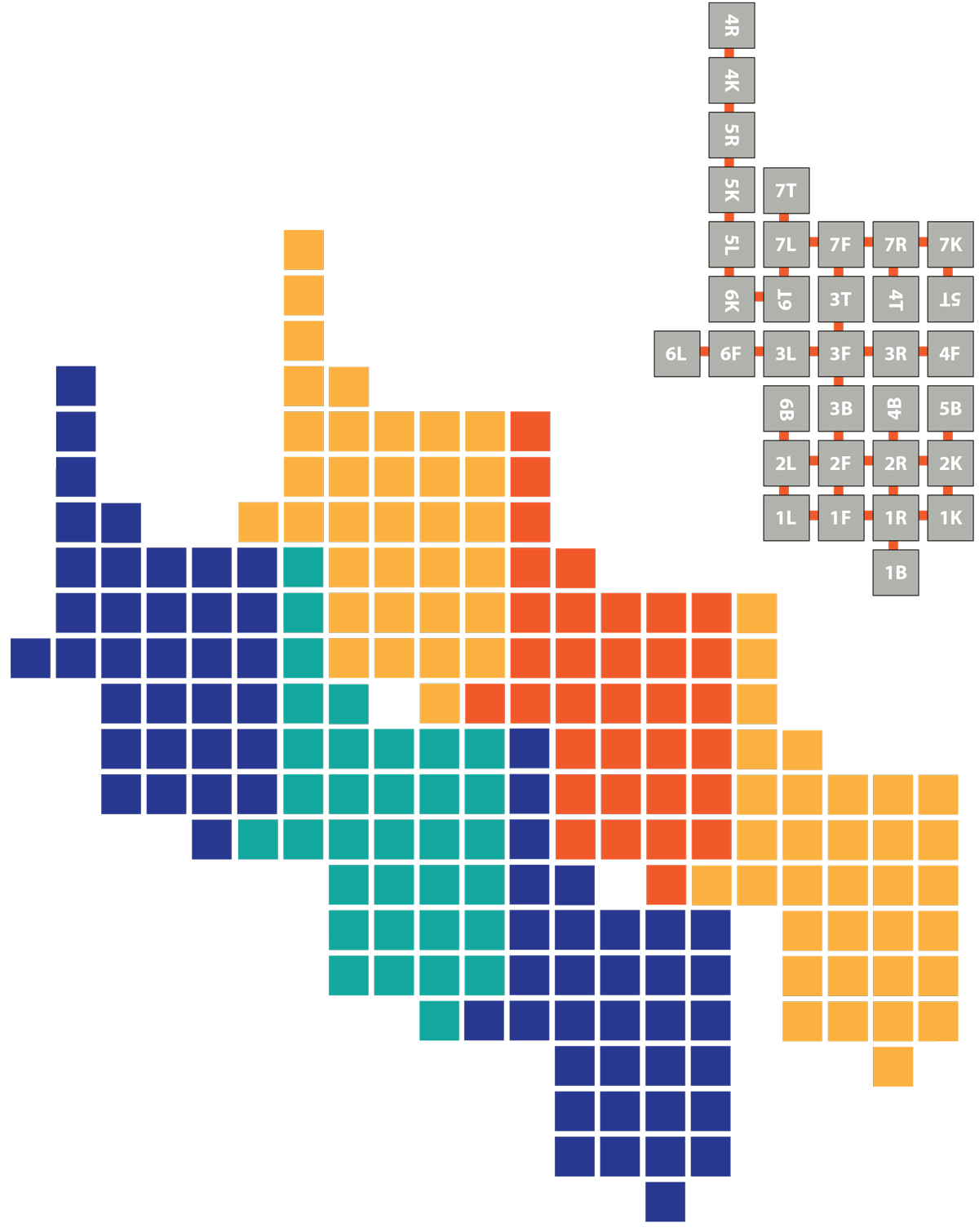}
\caption{An edge-unfolding of the Dali cross that nearly tiles the plane.
(See Figure~\protect\figref{Cross1_3D} for labels.)}
\figlab{hypercrossUnfolding}
\end{figure}

\section{Open Problems}
\seclab{Open}
\begin{enumerate}
\squeezelist
\item Is the $5$-dimensional cube in $\R^5$ a dimension-descending tiler?
\item What are good heuristics to test if the remaining $257$\footnote{
$261-4$, because $4$ are known to tile:
Figures~\protect\figref{L1_3D}, \protect\figref{Stadnicki}, \protect\figref{Cross1_3D}, \protect\figref{TTunfolding}.}
hypercube unfoldings tile $\R^3$?
\item Can any of the hypercube unfoldings be proved \underline{not} to tile $\R^3$?
\item Does the Dali cross have an unfolding that tiles $\R^2$?
\end{enumerate}

\paragraph{Addendum.}
We learned after posting this note that polyhedra
that have an edge-unfolding that tiles the plane
are called \emph{tessellation polyhedra} in~\cite{Stefan}.

\bibliographystyle{alpha}
\bibliography{CrossTiling}

\end{document}